\documentclass{moriond}
\usepackage{mathtools}

\DeclarePairedDelimiter\ket{\lvert}{\rangle}
\DeclarePairedDelimiterX\braket[2]{\langle}{\rangle}{#1\delimsize\vert #2}

\def\Journal#1#2#3#4{{#1} {\bf #2}, #3 (#4)}

\def\PRL{\em Phys. Rev. Lett.}
\def\PRD{{\em Phys. Rev.} D}

\def\be{\begin{equation}}
\def\ee{\end{equation}}
\def\bea{\begin{eqnarray*}}
\def\eea{\end{eqnarray*}}

\newcommand{\Photo}{\includegraphics[height=30mm]{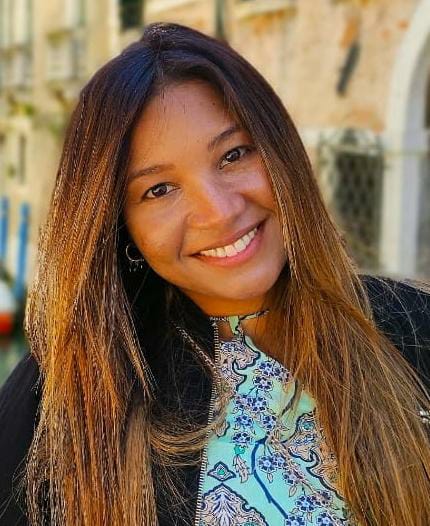}}
\begin{document}
\vspace*{4cm}

\title{Mixing and CP Violation in beauty and charm at LHCb}

\author{Laís Soares Lavra on behalf of the LHCb collaboration }

\address{Université Clermont Auvergne, Laboratoire de Physique de Clermont, CNRS/IN2P3,\\
Clermont-Ferrand, France}

\maketitle\abstracts{
This document presents eight recent and new  analyses from the LHCb experiment covering results on the CKM angle $\gamma$ and precision measurements of the charm mixing parameters. In addition, the latest {\em CP} violation measurements in charm and beauty decays are reported.}

\section{CKM angle $\gamma$}\label{sec:gamma}
The angle $\gamma \equiv \arg(-V_{ud}V^{*}_{ub}/V_{cd}V^{*}_{cb})$
of the CKM unitarity triangle can be measured directly using tree-level $b$-hadron decays. These measurements are obtained by exploiting the interference between $b\rightarrow u$ and $b\rightarrow c$ transitions in decays of $B^{\pm}\rightarrow Dh^{\pm}$, where $h=K,\pi$.\footnote{Hereinafter $h$ refer to either a charged kaon or a charged pion.} Since $\gamma$ is shared by all such decays, the best precision is obtained through the combination of measurements from several decay modes. 


\subsection{Update of the LHCb combination of the CKM angle $\gamma$}\label{subsec:combogamma}

Using LHCb data collected during the first two runs of the LHC,\footnote{First (second) run refers to the data collected in 2011-2012 at $\sqrt{s}$= 7-8 TeV (2015-2018 at $\sqrt{s}$= 13-14 TeV).} a combination of $\gamma$ measurements is presented using several $B\rightarrow Dh$ decay modes and LHCb results from different $D$-decay modes for the first time \cite{gammacombo}. Alongside $\gamma$, charm mixing and {\em CP} violation parameters are simultaneously determined in the combination. The results are obtained using frequentist treatment, and auxiliary inputs from other experiments. The combination yields $\gamma=(65.4^{+3.8}_{-4.2})^{\circ}$, and provides the best precision measurement from a single experiment to date. 

\subsection{Constraints on the CKM angle $\gamma$ from $B^{\pm}\rightarrow Dh^{\pm}$ decays }\label{subsec:latestgamma}
In this analysis \cite{constraintgamma}, $B^{\pm}\rightarrow Dh^{\pm}$ decays are studied using $D\rightarrow {K^{\pm}}{\pi^{\mp}}\pi^{0}$, $D\rightarrow K^{\pm}K^{\mp}\pi^{0}$, $D\rightarrow \pi^{\pm}\pi^{\mp}\pi^{0}$ and $D\rightarrow {\pi^{\pm}}{K^{\mp}}\pi^{0}$ final states. Measurements of $\gamma$ and related parameters $r_{B}$ (the magnitude of the ratio of amplitudes of the interfering $B$ decays) and $\delta_{B}$ (the strong phase between them) are obtained using 9 fb$^{-1}$ of data collected with the LHCb detector. The {\em CP} observables are determined through a simultaneous mass fit to all decay modes. The most suppressed decay mode $B^{\pm}\rightarrow [\pi^{\pm}K^{\mp}\pi^{0}]_{D}K^{\pm}$ is observed for the first time with a significance of 7.8 standard deviations, corresponding to 155 $\pm$ 19 signal events. The fit results and external inputs are used to interpret the results in terms of $\gamma$, $r_{B}$, $\delta_{B}$ and confidence intervals are evaluated using the profile likelihood method. The results are $\gamma=(56.4^{+24}_{-19})^{\circ}$, $\delta_{B}=(122^{+19}_{-23})^{\circ}$ and $r_{B}=(9.3^{+1.0}_{-0.9})\times 10^{-2}$. These measurements shall be used in future combinations to constrain the angle $\gamma$.

\section{CP violation and mixing in charm}
Charm meson oscillations are characterised by the mixing parameters, $x=\Delta{M}/{\Gamma}$ and 
$y=\Delta{\Gamma}/{(2\Gamma)}$, where $\Delta{M}$ and $\Delta{\Gamma}$ are the mass and decay width difference of the two mass eigenstates $\ket*{D_{1,2}}= p\ket*{D^{0}} \pm q\ket*{\bar{D}^{0}}$, and $\Gamma$ is the average decay width. {\em CP} violation in the mixing can occur if $|q/p|\neq1$ or in the interference between mixing and decay if $\phi_{f}=\arg(q\bar{{A}}_f/p{A}_f)\neq$ 0 or $\pi$ \footnote{${A}_f$($\bar{{A}}_f$) is the amplitude of the decay process $D^{0}(\bar{D}^{0})\rightarrow f$}. Flavour oscillation and {\em CP} violation in charm are heavily suppressed in the SM and expected to be very small effects, and precise determination of related parameters constitutes an important test to SM predictions.

\subsection{First observation of the mass difference between neutral charm-meson eigenstates}

A precise measurement of mixing and {\em CP}-violation parameters in $D^{0}\rightarrow K_{S}^{0}\pi^{+}\pi^{-}$ decay
is performed using 5.4 fb$^{-1}$ data sample collected by LHCb from 2016 to 2018~\cite{massdifference}. This analysis is Dalitz-plot and time dependent and the mixing parameters are determined using the {\em bin-flip} method.\cite{binflip} The results obtained for $x$, $y$, $|q/p|$ and $\phi$ are 
\begin{align*}
    x=(3.98^{+0.56}_{-0.54})\times10^{-3}, 
    &&
    y=(4.6^{+1.5}_{-1.4})\times 10^{-3}, 
    &&
    |q/p|=0.996\pm0.052,
    &&
    \phi=(0.056^{+0.047}_{-0.051})\,{\rm rad}.
\end{align*}

The value of the mixing parameter $x$ is found to be incompatible with zero with a significance greater than 7 standard deviations. This is the first observation of the mass difference of the charm neutral-meson eigenstates. All results are consistent with {\em CP} symmetry and improve the knowledge of mixing-induced {\em CPV} in the charm sector.  

\subsection{Precision measurement of the ${y}_{CP} - y_{CP}^{K\pi}$ parameter using two-body $D^{0}$ decays}

Using a 6 fb$^{-1}$ data sample collected by LHCb from 2015-2018, the charm mixing parameter ${y}_{CP}-y_{CP}^{K\pi}$~\footnote{ In previous publications the measured ${y}_{CP}-y_{CP}^{K\pi}$ was assumed equivalent to $y_{CP}$\cite{ycpkpi}, where $y_{CP}\approx y$ in the limit of {\em CP} conservation.} is precisely determined using the $D^{0}\rightarrow f$ ($f=K^-K^+$ or $\pi^-\pi^+$) and $D^{0}\rightarrow K^{-}\pi^{+}$ decays \cite{ycpmeasurement}. To determine the $y_{CP}-y^{K\pi}_{CP}$ parameter, a weighted average of statistically independent measurements of $y^{KK}_{CP}-y^{K\pi}_{CP}$ and $y^{\pi\pi}_{CP}-y^{K\pi}_{CP}$ is performed. These parameters are accessed  with an exponential fit to the signal yield ratios of $D^{0}\rightarrow f$ over $D^{0}\rightarrow K^{-}\pi^{+}$ as a function of the decay time $t$, 
\begin{equation}
\label{eq:Rt}
    R^{f}(t) = \frac{N(D^{0}\rightarrow f, t)}{N(D^{0}\rightarrow K^{-}\pi^{+}, t)} \propto {e}^{-({y^{f}_{CP} - y^{K\pi}_{CP}})t/\tau_{D^{0}}} \times \frac{\varepsilon(f,t)}{\varepsilon(K^{-}\pi^{+},t)},
\end{equation}
 where $\tau_{D^{0}}$ is the $D^{0}$ lifetime and $\varepsilon(h^-h^+,t)$ is the time-dependent efficiency. Since the efficiencies differ for each decay mode, a procedure based on matching the kinematic phase space and correcting the difference of detection efficiencies is employed to equalise the efficiencies so that they cancel in the ratio. The results of the fit to the corresponding $R^{f}(t)$ are
 \bea
 y^{\pi\pi}_{CP}-y^{K\pi}_{CP}=(6.57\pm0.53\pm0.16)\times 10^{-3}, &
y^{KK}_{CP}-y^{K\pi}_{CP}=(7.08\pm0.30\pm0.14)\times10^{-3},
 \eea
where the first uncertainty is statistical and the second systematic. Combining these two measurements yields $y_{CP}-y^{K\pi}_{CP}= (6.96\pm0.26\pm0.13)\times10^{-3}$, which is compatible with the world average and more precise by a factor of four. 
\subsection{Measurement of $\mathit{CP}$ asymmetries in $D^{+}_{(s)}\rightarrow \eta\pi^{+}$ and $D^{+}_{(s)}\rightarrow \eta^{'}\pi^{+}$ decays}
Searches for direct {\em CP} asymmetries in $D^{+}_{(s)}\rightarrow \eta\pi^{+}$ and $D^{+}_{(s)}\rightarrow \eta'\pi^{+}$ decays are performed using 6 fb$^{-1}$ LHCb dataset from 2015-2018, with $\eta^{(')}$ mesons reconstructed in the final state $\gamma \pi^{+}\pi^{-}$~\cite{Detapi}. The {\em CP} asymmetry $\mathcal{A}^{CP}$ of each decay mode is obtained by subtracting from the raw asymmetry the production and detection asymmetries, determined using the control channels $D^{+}_{(s)}\rightarrow \phi\pi^{+}$. A simultaneous likelihood fit to the invariant mass distribution of the data is performed for $D^{\pm}_{(s)}$ candidates in intervals of $\eta^{(')}$ mass to determine the raw asymmetries. The measured {\em CP} asymmetries are
\begin{eqnarray*}
 \mathcal{A}^{CP}_{D^{+}\rightarrow \eta \pi^{+}}=(0.34\pm0.66\pm0.16\pm 0.05)\%, 
  &
  \mathcal{A}^{CP}_{D_{s}^{+}\rightarrow \eta \pi^{+}}=(0.32\pm 0.51 \pm 0.12)\%,
    \\
  \mathcal{A}^{CP}_{D^{+}\rightarrow \eta{'} \pi^{+}}=(0.49\pm 0.18 \pm 0.06 \pm 0.05 ) \%,
    & 
  \mathcal{A}^{CP}_{D_{s}^{+}\rightarrow \eta' \pi^{+}}=(0.01\pm 0.12 \pm 0.08) \%,
\end{eqnarray*}
where the first uncertainty is statistical, the second systematic and, for the $D^+$ modes, the last comes from the uncertainty on the {\em CP} asymmetry of the $D^{+}\rightarrow \phi\pi^{+}$ control channel. All the results are consistent with {\em CP} symmetry, and the measurements are the most precise to date for the $D^{+}\rightarrow \eta \pi^{+}$, $D^{+}\rightarrow \eta{'} \pi^{+}$ and $D_{s}^{+}\rightarrow \eta' \pi^{+}$ channels. 

\section{CP violation in charmless b-meson decays}
Charmless {\em b}-meson decays provide an interesting environment to observe {\em CP} violation effects since these decays have contributions
from both penguin and tree-level processes with amplitudes of similar size. In addition, in multi-body decays the different intermediate states can interfere with each other, producing large strong-phase differences, which can lead to enhanced {\em CP} violation in certain regions of phase space. 
\subsection{Direct $\mathit{CP}$ violation in charmless three-body decays of $B^{\pm}$ mesons}
\begin{figure}
\begin{minipage}{0.5\linewidth}
\centerline{\includegraphics[width=0.73\linewidth]{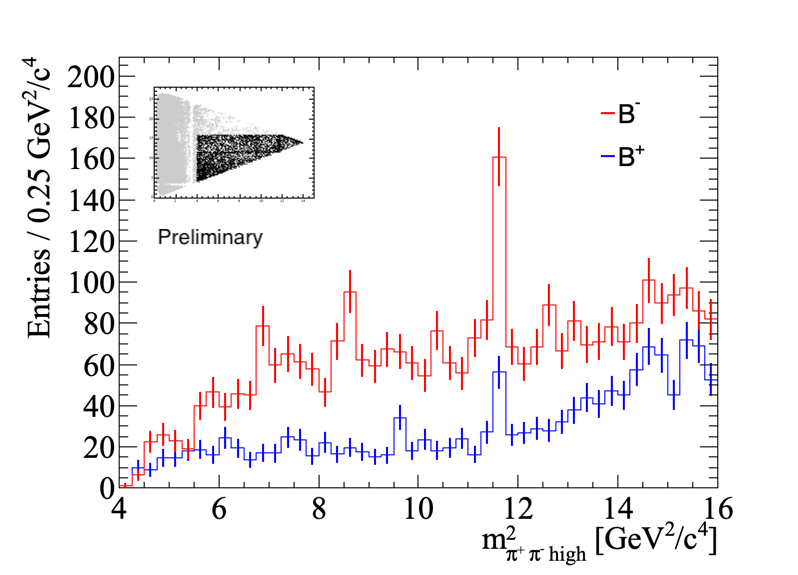}}
\end{minipage}
\begin{minipage}{0.5\linewidth}
\centerline{\includegraphics[width=0.97\linewidth]{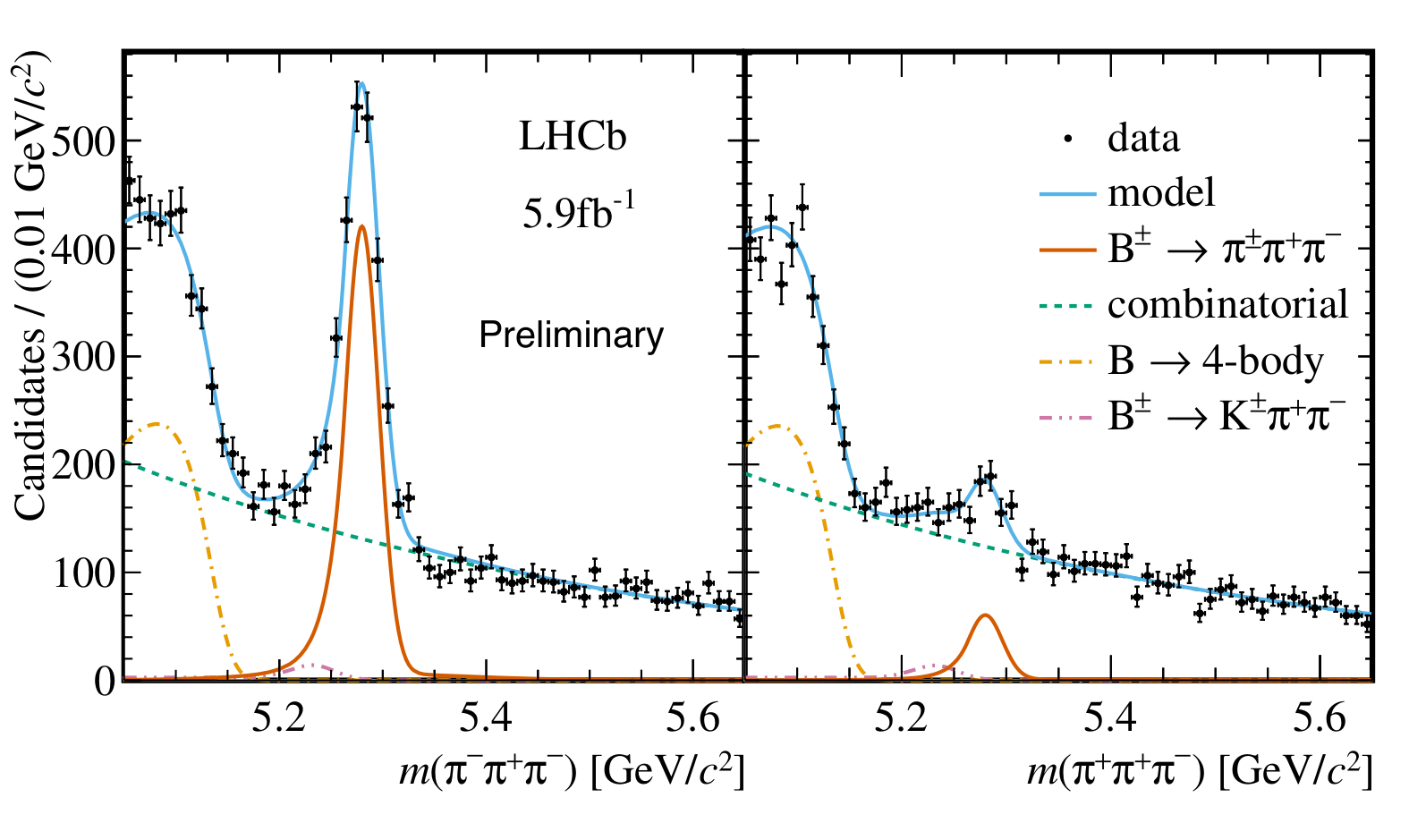}}
\end{minipage}
\caption{The $B^{\pm}\rightarrow \pi^{\pm}\pi^{+}\pi^{-}$: (left) $m^2(\pi^{+}\pi^{-})_{\text{high}}$ projection for the region $4<m^2(\pi^{+}\pi^{-})_{\text{high}}<16$ GeV$^{2}/$c$^4$ and $4<m^2(\pi^{+}\pi^{-})_{\text{low}}<15$ GeV$^{2}/$c$^4$  with (right) mass fits for this region ($B^-$ on the left, $B^+$ on the right).}
\label{fig:3pi_Xic}
\end{figure}

Measurements of direct $\mathit{CP}$ violation in $B^{\pm}\rightarrow K^{\pm}\pi^{+}\pi^{-}$, $B^{\pm}\rightarrow K^{\pm}K^{+}K^{-}$, $B^{\pm}\rightarrow \pi^{\pm}\pi^{+}\pi^{-}$ and $B^{\pm}\rightarrow \pi^{\pm}K^{+}K^{-}$ decays are performed using 5.9 fb$^{-1}$ data sample collected by LHCb from 2015-2018~\cite{b3h}. The {\em CP} asymmetry $\mathcal{A}^{CP}$ of each decay mode are accessed directly from the raw asymmetry, corrected for nuisance asymmetries. The raw asymmetries are obtained from the phase-space efficiency corrected yields of $B^{+}$ and $B^{-}$ fitted simultaneously. Then the {\em CP} asymmetries are obtained by subtracting the production asymmetry, calculated using the control mode $B^{\pm}\rightarrow J/\psi K^{\pm}$. The results for the phase-space integrated {\em CP} asymmetries are
\begin{eqnarray*}
\mathcal{A}^{CP}_{B^{\pm}\rightarrow K^{\pm}\pi^{+}\pi^{-}}=(+1.1\pm0.2\pm0.3\pm0.3)\%, 
&
\mathcal{A}^{CP}_{B^{\pm}\rightarrow K^{\pm}K^{+}K^{-}}=(-3.7\pm0.2\pm0.2\pm0.3)\%,
\\
\mathcal{A}^{CP}_{B^{\pm}\rightarrow \pi^{\pm}\pi^{+}\pi^{-}}=(+8.0\pm0.4\pm0.3\pm0.3)\%, &
\mathcal{A}^{CP}_{B^{\pm}\rightarrow \pi^{\pm}K^{+}K^{-}}=(-11.4\pm0.7\pm0.3\pm0.3)\%, 
\end{eqnarray*}
where the first uncertainty is statistical, second systematic, and third uncertainty due to the {\em CP} asymmetry of the $B^{\pm}\rightarrow J/\psi K^{\pm}$ control channel. For $B^{\pm}\rightarrow \pi^{\pm}\pi^{+}\pi^{-}$ and $B^{\pm}\rightarrow K^{\pm}K^{+}K^{-}$ decays, {\em CP} violation is observed for the first time with significance of 14.1$\sigma$ and 8.5$\sigma$, respectively. For $B^{\pm}\rightarrow \pi^{\pm}K^{+}K^{-}$ the observed {\em CP} violation is compatible and more precise than the previous measurement while the result for the $B^{\pm}\rightarrow K^{\pm}\pi^{+}\pi^{-}$ decay is consistent with {\em CP} conservation. The {\em CP} asymmetry distributions across the Dalitz Plot of these four modes are investigated using the \textit{Mirandizing} approach~\cite{mirandizing}. Nine different regions of the Dalitz plot are studied among the four $B^{\pm}\rightarrow h^{\pm}h^{+}h^{-}$ decays. The local {\em CP} asymmetries are measured from fits
to the invariant mass of candidates in the regions studied, following the same procedure as for integrated measurements. In Fig.~\ref{fig:3pi_Xic} shows the results for a selected region in $B^{\pm}\rightarrow \pi^{\pm}\pi^{+}\pi^{-}$ decay. The projection
onto $m^2(\pi^{+}\pi^{-})_{\text{high}}$ reveals
an indication of the $\chi_{c0}$(1P) resonance for the first time. The local {\em CP} asymmetry measured in this region is $\mathcal{A}^{CP} =({+74.5}\pm2.7\pm1.8\pm0.3)\%$, which is the highest {\em CP} asymmetry ever measured to date.
In addition, significant localised {\em CP} asymmetries in the $\pi\pi \rightarrow KK$ rescattering region are observed in the four analysed channels.

\subsection{Search for direct $\mathit{CP}$ violation in charged charmless $B\rightarrow PV$ decays}\label{sec:b3h}
Measurements of {\em CP} asymmetries are reported in charmless $B$ decays to a pseudo-scalar and a vector resonance ($B\rightarrow PV$) using 5.9 fb$^{-1}$ data sample collected by LHCb from 2015-2018~\cite{bpv}. A new model-independent method based on the angular distributions is employed to measure {\em CP} asymmetries in low mass and narrow vector resonances~\cite{bpvmodel}. Five different $B\rightarrow PV$ decays from $B^{\pm}\rightarrow h^{\pm}h^{+}h^{-}$ final states are analysed. For the  $B^\pm \to\rho(770)^{0}K^\pm$ decay a {\em CP} asymmetry of $\mathcal{A}^{CP}=(+15.0\pm 1.9 \pm 1.1)\%$ is observed for the first time with a significance of 6.8$\sigma$. For the other modes, results are consistent with {\em CP} symmetry. 

\subsection{Search for $CP$ violation using $\hat{T}$-odd correlations in $B^{0}\rightarrow p \bar{p}K^{+}\pi^{-}$ decays}
Using 8.4 fb$^{-1}$ LHCb dataset from 2011-2012 and 2016-2018, a search for {\em CP} and {\em P} violation in charmless four-body $B^{0}\rightarrow p \bar{p}K^{+}\pi^{-}$ decays is performed with observables constructed using triple-product asymmetries techniques~\cite{baryonCPV,tpa}. These observables are measured both globally and in regions of the phase-space, where their deviation from zero is probed. The signal yields and the observables are obtained from a simultaneous fit to $p \bar{p}K^{+}\pi^{-}$ invariant mass distributions. No evidence of {\em CP} and {\em P} asymmetries are found in integrated measurements. Results in regions of the phase are consistent with {\em CP} symmetry and local {\em P}-violations with significance greater than 5.8$\sigma$ are observed. 

\end{document}